\documentclass[amsmath,floatfix,twocolumn,superscriptaddress]{revtex4-1}
\usepackage{amssymb,amsmath}
\usepackage{graphicx,array}
\usepackage{dcolumn}
\usepackage{psfrag}
\usepackage{bm}
\usepackage{color}

\begin{document}
\title{Colossal barocaloric effects in the complex hydride Li$_{2}$B$_{12}$H$_{12}$} 

\author{Kartik Sau}
\affiliation{Mathematics for Advanced Materials - Open Innovation Laboratory (MathAM-OIL), 
             National Institute of Advanced Industrial Science and Technology (AIST), c/o 
	     Advanced Institute of Material Research (AIMR), Tohoku University, Sendai 980-8577, Japan}

\author{Tamio Ikeshoji}
\affiliation{Mathematics for Advanced Materials - Open Innovation Laboratory (MathAM-OIL), 
             National Institute of Advanced Industrial Science and Technology (AIST), c/o 
             Advanced Institute of Material Research (AIMR), Tohoku University, Sendai 980-8577, Japan}

\author{Shigeyuki Takagi}
\affiliation{Institute for Materials Research, Tohoku University, Sendai 980-8577, Japan}

\author{Shin-ichi Orimo}
\affiliation{Institute for Materials Research, Tohoku University, Sendai 980-8577, Japan}
\affiliation{Advanced Institute for Materials Research, Tohoku University, Sendai 980-8577, Japan}

\author{Daniel Errandonea}
\affiliation{Departament de F\'isica Aplicada, Institut de Ci\`encia de Materials, MALTA Consolider Team, Universitat de Val\`encia, 
             Edifici d'Investigaci\'o, Burjassot 46100, Spain}

\author{Dewei Chu}
\affiliation{School of Materials Science and Engineering, UNSW Sydney, NSW 2052, Australia}

\author{Claudio Cazorla}
\thanks{Corresponding Author}
\affiliation{School of Materials Science and Engineering, UNSW Sydney, NSW 2052, Australia}

\maketitle

{\bf Traditional refrigeration technologies based on compression cycles of greenhouse gases 
pose serious threats to the environment and cannot be downscaled to electronic device dimensions.
Solid-state cooling exploits the thermal response of caloric materials to external fields 
and represents a promising alternative to current refrigeration methods. However, most of 
the caloric materials known to date present relatively small adiabatic temperature changes
($|\Delta T| \sim 1$~K) and/or limiting irreversibility issues resulting from significant 
phase-transition hysteresis. Here, we predict the existence of colossal barocaloric effects 
(isothermal entropy changes of $|\Delta S| \sim 100$~JK$^{-1}$kg$^{-1}$) in the energy material 
Li$_{2}$B$_{12}$H$_{12}$ by means of molecular dynamics simulations. Specifically, we estimate 
$|\Delta S| = 387$~JK$^{-1}$kg$^{-1}$ and $|\Delta T| = 26$~K for an applied pressure of $P = 
0.4$~GPa at $T = 475$~K. The disclosed colossal barocaloric effects are originated by an 
order-disorder phase transformation that exhibits a fair degree of reversibility and involves 
coexisting Li$^{+}$ diffusion and (BH)$_{12}^{-2}$ reorientational motion at high temperatures.}
\\

Solid-state cooling is an environmentally friendly, energy efficient, and highly scalable technology 
that can solve most of the problems associated with conventional refrigeration methods based on compression 
cycles of greenhouse gases (i.e., environmental harm and lack of downsize scaling). Upon application 
of magnetic, electric or stress fields good caloric materials undergo noticeable temperature changes 
($|\Delta T| \sim 1$--$10$~K) as a result of induced phase transformations that involve large entropy 
variations ($|\Delta S| \sim 10$--$100$~JK$^{-1}$kg$^{-1}$) \cite{manosa17,moya14,cazorla19}. 
Solid-state cooling capitalizes on such caloric effects to engineer refrigeration cycles. From a 
performance point of view, that is, largest $|\Delta T|$ and $|\Delta S|$ (although these are not 
the only parameters defining cooling efficiency \cite{lloveras20}), barocaloric effects driven by small 
hydrostatic pressure shifts appear to be the most promising \cite{manosa17,moya14,cazorla19}. 

\begin{figure}[t]
\centerline
        {\includegraphics[width=1.00\linewidth]{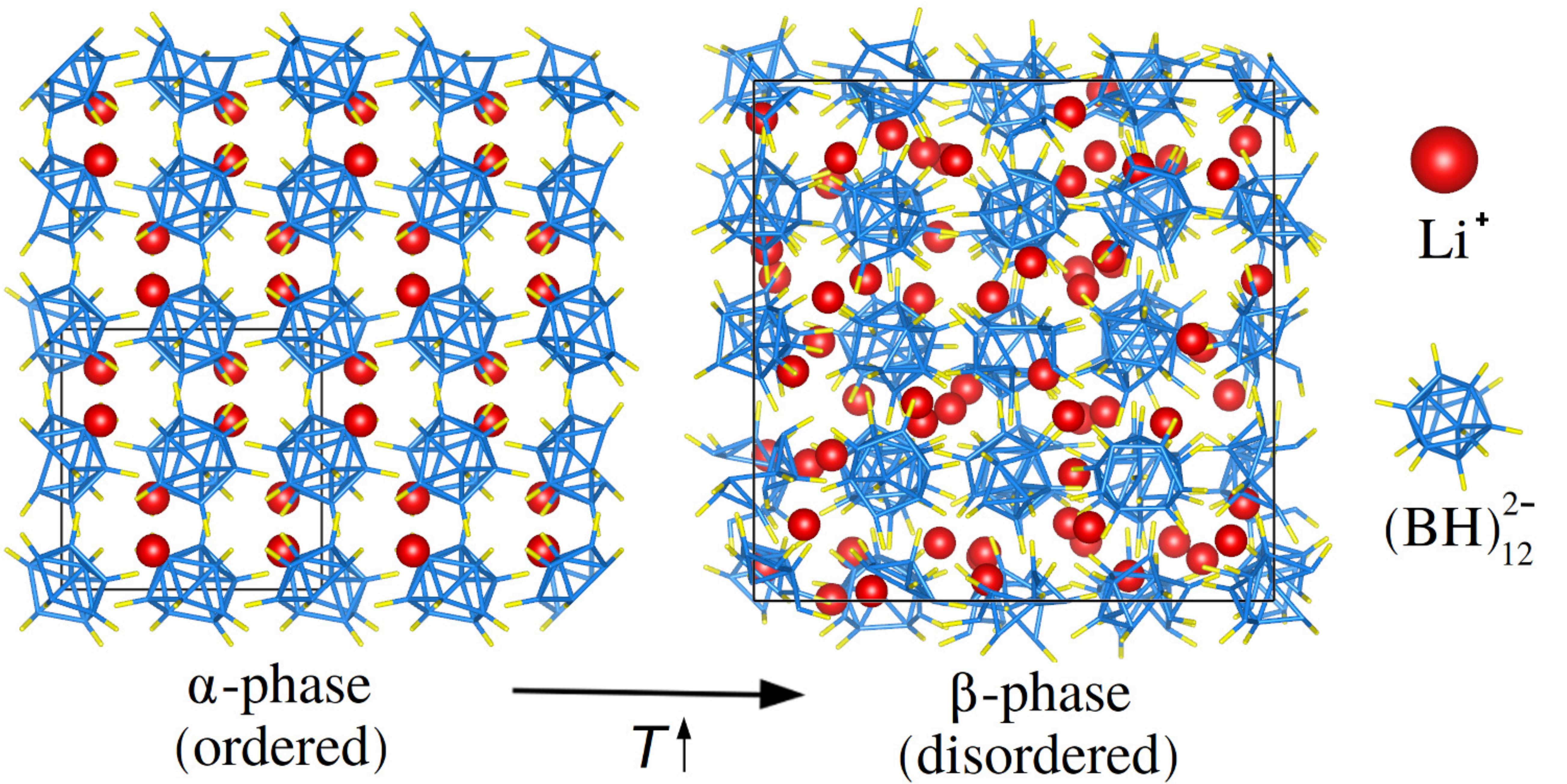}}
        \caption{Low-$T$ (ordered) and high-$T$ (disordered) phases of bulk Li$_{2}$B$_{12}$H$_{12}$.
        The low-$T$ phase ($\alpha$) presents cubic symmetry and space group $Pa\overline{3}$ \cite{her08}.
        In the high-$T$ phase ($\beta$), cubic symmetry is preserved but the Li$^{+}$ ions are highly
	mobile and the (BH)$_{12}^{-2}$ icosahedra present reorientational disorder \cite{paskevicius13}.
        The $T$-induced $\alpha \to \beta$ phase transition is an order-disorder isosymmetric transformation
        \cite{paskevicius13}. Li, B, and H ions are represented with red, blue, and yellow colours,
        respectively.}
\label{fig1}
\end{figure}

Recently, colossal barocaloric effects (defined here as $|\Delta S| \sim 100$~JK$^{-1}$kg$^{-1}$) have 
been measured in two different families of materials that display intriguing order-disorder phase transitions 
\cite{cazorla17a,li19,lloveras19}. First, giant barocaloric effects have been theoretically predicted 
\cite{cazorla17b} and experimentally observed in the archetypal superionic compound AgI \cite{cazorla17a}. 
AgI exhibits a first-order normal (low-entropy) to superionic (high-entropy) phase transition that responds 
to both temperature and pressure \cite{sagotra17} and which involves the presence of highly mobile silver 
ions in the high--$T$ superionic state \cite{hull04}. The entropy changes estimated for other normal to
superionic phase transitions in general are large as well \cite{cazorla16,cazorla18,min20,cazorla19a}. And second, 
colossal barocaloric effects have been reported for the molecular solid neopentylglycol \cite{li19,lloveras19}, 
(CH$_{3}$)$_{2}$C(CH$_{2}$OH)$_{2}$, and other plastic crystals \cite{lloveras20}. In these solids molecules 
reorient almost freely around their centers of mass, which remain localized at well-defined lattice positions. 
Molecular rotations lead to orientational disorder, which renders high entropy. By using hydrostatic pressure, 
it is possible to block such molecular reorientational motion and thus induce a fully ordered state characterized 
by low entropy \cite{cazorla19a}. The barocaloric effects resulting from this class of order-disorder phase 
transition are huge and comparable in magnitude to those achieved in conventional refrigerators with environmentally 
harmful fluids \cite{lloveras20,li19,lloveras19}. 

Here, we report the prediction of colossal barocaloric effects ($|\Delta S| \sim 100$~JK$^{-1}$kg$^{-1}$) 
in the energy material Li$_{2}$B$_{12}$H$_{12}$ (LBH), a complex hydride that is already known from the 
fields of hydrogen storage \cite{her08,lai19,shevlin12} and solid-state batteries \cite{paskevicius13,luo20,mohtadi16}. 
By using molecular dynamics simulations, we identify a pressure-induced isothermal entropy change of $|\Delta S| 
= 387$~JK$^{-1}$kg$^{-1}$ and adiabatic temperature change of $|\Delta T| = 26$~K at $T = 475$~K. These 
colossal entropy and temperature changes are driven by moderate hydrostatic pressure shifts of $P = 0.4$~GPa, 
thus yielding huge barocaloric strengths of $|\Delta S| / P = 968$~JK$^{-1}$kg$^{-1}$GPa$^{-1}$ and $|\Delta T| 
/ P = 65$~K~GPa$^{-1}$. The colossal barocaloric effects disclosed in bulk LBH are originated by simultaneous 
$P$-driven frustration and activation of Li$^{+}$ diffusion and (BH)$_{12}^{-2}$ icosahedra reorientational 
motion. Thus, alkali-metal complex borohydrides ($A_{2}$B$_{12}$H$_{12}$, $A =$ Li, Na, K, Cs \cite{udovic14,udovic20}) 
emerge as a promising new family of barocaloric materials in which the salient phase-transition features of 
fast-ion conductors and plastic crystals coexist.  

\begin{figure*}[t]
\centerline
        {\includegraphics[width=1.00\linewidth]{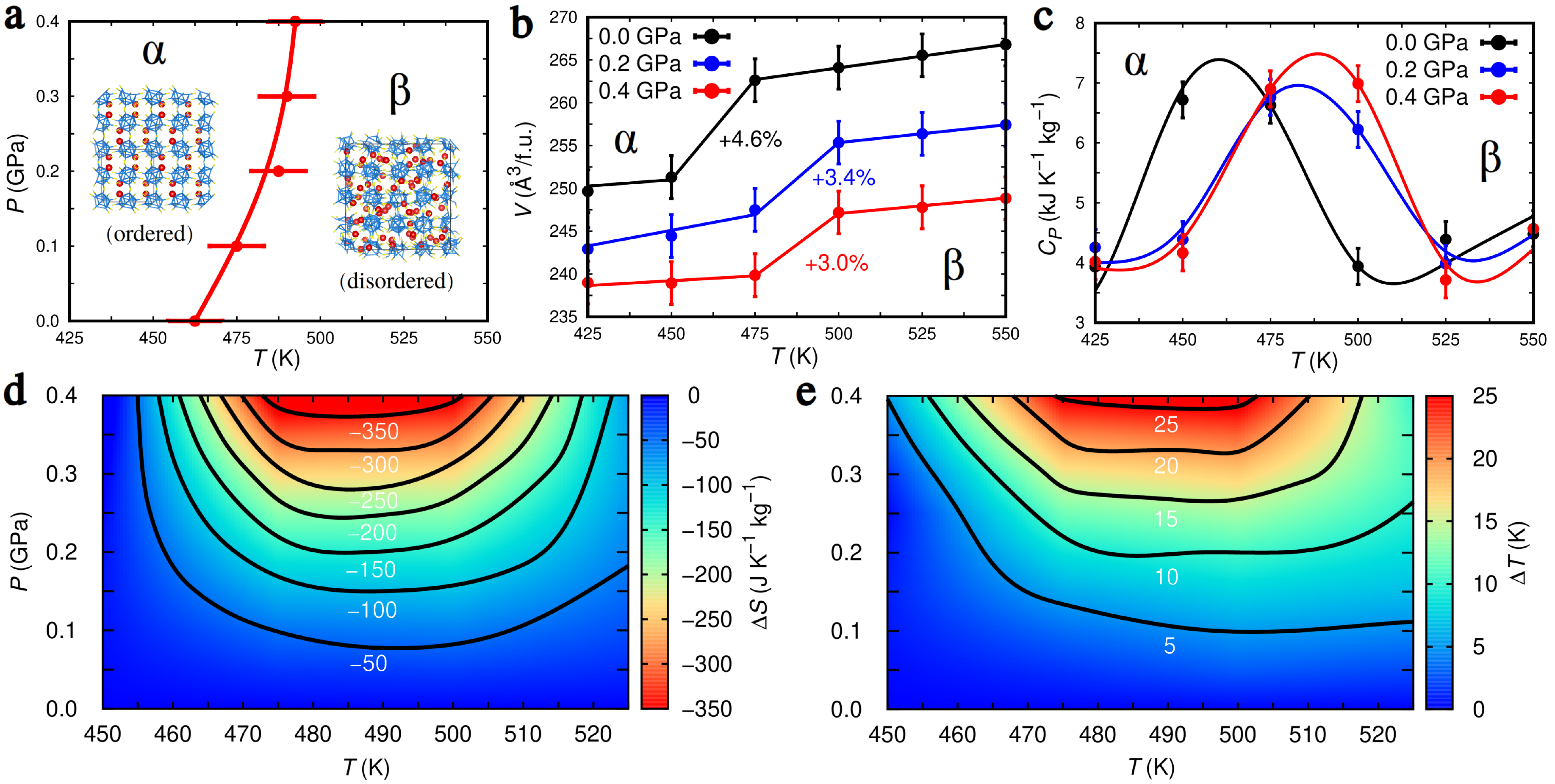}}
        \caption{Influence of pressure on the $T$-induced $\alpha \to \beta$ phase transition occurring in bulk
        Li$_{2}$B$_{12}$H$_{12}$ and the resulting barocaloric effects. {\bf a}~Estimated $P$-$T$ phase
        boundary separating the stability regions of the $\alpha$ and $\beta$ phases. {\bf b}~Volume change
        estimated for the $T$-induced $\alpha \to \beta$ phase transition at different pressures. {\bf c}~Heat
        capacity of bulk Li$_{2}$B$_{12}$H$_{12}$ expressed as a function of temperature and pressure.
        {\bf d}~Isothermal entropy and {\bf e}~adiabatic temperature changes associated with the barocaloric
        response of bulk Li$_{2}$B$_{12}$H$_{12}$ expressed as a function of applied pressure and temperature. 
	Solid black lines represent isovalue curves.}
\label{fig2}
\end{figure*}

\begin{figure*}[t]
\centerline
        {\includegraphics[width=1.00\linewidth]{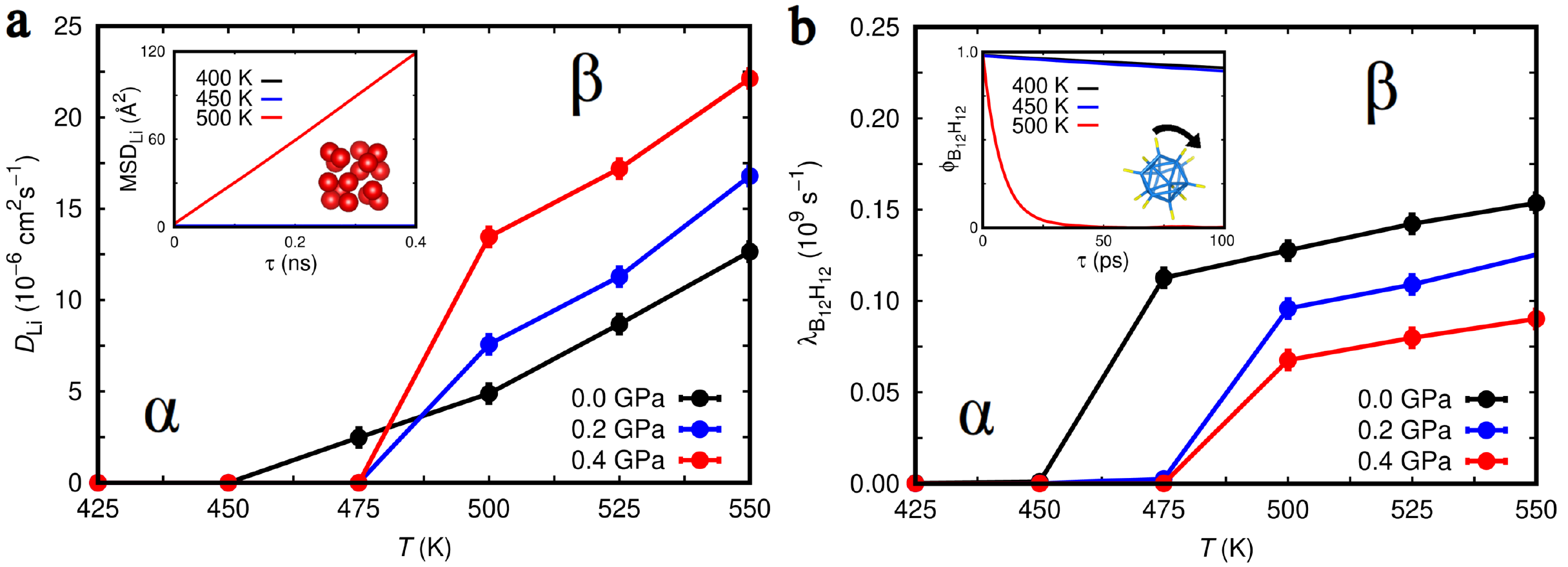}}
        \caption{Order parameter changes associated with the $T$-induced $\alpha \to \beta$ phase transition
        occurring in bulk Li$_{2}$B$_{12}$H$_{12}$ at different pressures. {\bf a}~Estimated lithium ion diffusion
        coefficient, $D_{\rm Li}$, expressed as a function of temperature and pressure. The inset shows
        the Li mean-squared displacement (MSD$_{\rm Li}$) data employed for the calculation of $D_{\rm Li}$ at
        zero pressure (Methods). {\bf b}~Estimated (BH)$_{12}^{-2}$ icosahedra reorientational rate, $\lambda_{\rm B_{12}H_{12}}$,
        expressed as a function of temperature and pressure. The inset shows the (BH)$_{12}^{-2}$ icosahedra
        angular auto-correlation function ($\phi_{\rm B_{12}H_{12}}$) data employed for the calculation of
        $\lambda_{\rm B_{12}H_{12}}$ at zero pressure (Methods).}
        \label{fig3}
\end{figure*}

\section*{RESULTS}
At ambient conditions, lithium dodecahydrododecaborate (Li$_{2}$B$_{12}$H$_{12}$), LBH, presents an ordered 
cubic $Pa\overline{3}$ phase, referred to as $\alpha$ hereafter, which is characterized by Li$^{+}$ cations 
residing on near-trigonal-planar sites surrounded by three (BH)$_{12}^{-2}$ icosahedron anions. In turn, each 
(BH)$_{12}^{-2}$ anion resides in an octahedral cage surrounded by six Li$^{+}$ cations (Fig.\ref{fig1}a) 
\cite{her08}. A symmetry preserving order-disorder phase transition occurs at high temperatures ($\sim 500$~K) 
that stabilises a disordered state, referred to as $\beta$ hereafter, in which the Li$^{+}$ cations are mobile 
and the (BH)$_{12}^{-2}$ anions present reorientational motion (Fig.\ref{fig1}b) \cite{paskevicius13}. The 
relative volume expansion that has been experimentally measured for such an order-disorder phase transition is 
$\approx 8$\% \cite{paskevicius13}. This huge volume variation along with the accompanying, and pressumably 
also large, phase-transition entropy change could be propitious for barocaloric purposes if the involved phase 
transformation was responsive to moderate external pressures of $\sim 0.1$~GPa. To the best of our knowledge, 
this possibility has not been hitherto explored. We performed classical molecular dynamics (MD) simulations 
based on a recently proposed LBH force field \cite{sau19} to fill up such a knowledge gap (Methods and 
Supplementary Methods), which has clear implications for potential solid-state cooling applications.

Figure~\ref{fig2}a shows the $P$--$T$ phase diagram that we estimated for bulk LBH using atomistic MD simulations.
It was found that the temperature of the $\alpha \to \beta$ phase transition is certainly sensitive to external pressure. 
Specifically, the $dP / dT$ derivative of the corresponding phase boundary amounts to $\approx 0.008$~GPa~K$^{-1}$ 
at zero pressure and to $\approx 0.02$~GPa~K$^{-1}$ at $P = 0.2$~GPa. Likewise, the relative volume change ascribed 
to the $\alpha \to \beta$ transformation is, according to our simulations, $+4.6$\% at zero pressure and $+3.4$\% 
at $P = 0.2$~GPa (Fig.\ref{fig2}b). By using these thermodynamic data and the Clausius-Clapeyron relation \cite{moya14}, 
we roughly estimated an entropy change of $\Delta S \sim 300$~JK$^{-1}$kg$^{-1}$ for the order-disorder transition 
occurring in LBH at $P = 0.2$~GPa. In view of these promising barocaloric descriptor values, we proceeded to accurately 
calculate the barocaloric isothermal entropy and adiabatic temperature changes, $\Delta S$ and $\Delta T$, induced 
by pressures $0 \le P \le 0.4$~GPa. To this end, we followed the numerical protocols described in the Methods section, 
which essentially involve the determination of the volume and heat capacity of bulk LBH (Fig.\ref{fig2}c) as a 
function of pressure and temperature.

The results of our precise barocaloric calculations for temperatures and pressures in the intervals $450 \le T \le 
525$~K and $0 \le P \le 0.4$~GPa are shown in Figs.~\ref{fig2}d,e. The $\Delta S$ and $\Delta T$ values estimated 
for the $\alpha \to \beta$ transformation in fact render colossal barocaloric effects. For example, at $T = 490$~K 
and $P = 0.4$~GPa ($0.2$~GPa) we calculated an isothermal entropy change of $-365$~JK$^{-1}$kg$^{-1}$ 
($-135$~JK$^{-1}$kg$^{-1}$) and an adiabatic temperature change of $+27$~K ($+10$~K). The resulting barocaloric 
effects are direct, that is, $\Delta T > 0$, because the low-entropy ordered state is stabilized under pressure 
($\Delta S < 0$). A maximum $|\Delta S|$ value of $387$~JK$^{-1}$kg$^{-1}$ was found at $T = 475$~K and $P = 0.4$~GPa 
(Fig.\ref{fig2}d). For temperatures above $\approx 510$~K, we estimated noticeably smaller $|\Delta S|$ and $\Delta T$ 
values (e.g., $72$~JK$^{-1}$kg$^{-1}$ and $10$~K for $P = 0.4$~GPa at $T = 525$~K), a trend that we link to some 
anomalous pressure-induced ionic diffusion (explained below). In the Discussion section, we will compare the barocaloric 
performance of LBH with those of other well-known barocaloric materials. In what follows, the atomistic mechanisms 
leading to the extraordinary $\Delta S$ and $\Delta T$ results just reported are unravelled. 

There are two possible sources of large entropy variation in LBH, one stemming from the Li$^{+}$ ionic diffusion 
and the other from the (BH)$_{12}^{-2}$ icosahedra reorientational motion. When hydrostatic pressure is applied 
on the disordered $\beta$ phase at temperatures below $\approx 500$~K, both the ionic diffusion and molecular 
orientational disorder are reduced and thus the crystal entropy diminishes significantly. This conclusion is
straightforwardly deduced from the $P$-induced variation of the Li$^{+}$ diffusion coefficient, $D_{\rm Li}$, 
and reorientational (BH)$_{12}^{-2}$ frequency, $\lambda_{\rm B_{12}H_{12}}$, shown in Figs.\ref{fig3}a,b (Methods). 
For instance, at $T = 475$~K and zero pressure $D_{\rm Li}$ and $\lambda_{\rm B_{12}H_{12}}$ amount to $2.5 \cdot 
10^{-6}$~cm$^{2}$s$^{-1}$ and $1.2 \cdot 10^{8}$~s$^{-1}$, respectively, whereas at $P = 0.2$~GPa both quantities 
are practically zero (Fig.\ref{fig3}). The two resulting contributions to the system entropy variation are of the 
same sign and make $|\Delta S|$ huge. 

Which of these two $P$-induced order-restoring effects is most relevant for the barocaloric performance of bulk LBH? 
To answer this question, we performed constrained MD simulations in which we forced the Li$^{+}$ ions to remain localized 
around their equilibrium positions independently of temperature. This type of artificial condition in principle cannot 
be imposed in the experiments but can be easily enforced in the atomistic simulations. The $|\Delta S|$ values estimated 
in these constrained MD simulations were roughly half the value of the isothermal entropy changes obtained in the standard 
MD simulations. Therefore, we may conclude that at temperatures below $\approx 500$~K the pressure-induced entropy changes
stemming from the Li$^{+}$ ionic diffusion and (BH)$_{12}^{-2}$ icosahedra reorientational motion variations play both 
an equally important role in the global barocaloric response of LBH.

Figure~\ref{fig3}a shows that at $T \gtrsim 500$~K the Li$^{+}$ diffusion coefficient increases under increasing 
pressure. For example, at $T = 525$~K and zero pressure we estimate $D_{\rm Li} = 8.7 \cdot 10^{-6}$~cm$^{2}$s$^{-1}$ 
whereas at $P = 0.4$~GPa and the same temperature we obtain $17.2 \cdot 10^{-6}$~cm$^{2}$s$^{-1}$. This ionic 
diffusion behaviour is highly anomalous because hydrostatic compression typically hinders ionic transport  
\cite{sagotra17,hull04,cazorla16}. On the other hand, the reorientational motion of the (BH)$_{12}^{-2}$ 
icosahedra behaves quite normally, that is, decreases under pressure \cite{cazorla19,lloveras20}. For instance, at $T = 
525$~K and zero pressure we estimate $\lambda_{\rm B_{12}H_{12}} = 1.4 \cdot 10^{8}$~s$^{-1}$ whereas at $P = 0.4$~GPa 
and the same temperature we obtain $0.7 \cdot 10^{8}$~s$^{-1}$ (Fig.\ref{fig3}b). We hypothesize that the anomalous 
$P$-induced Li$^{+}$ diffusion behaviour observed in our MD simulations is due to the high anionic reorientational 
motion, which makes the (BH)$_{12}^{-2}$ centers of mass to fluctuate and partially block the ionic current channels 
\cite{skripov13}. Consistently, when the frequency of the (BH)$_{12}^{-2}$ rotations is reduced by effect of compression 
the ions can flow more easily throughout the crystal and Li$^{+}$ transport is enhanced. In this particular $P$--$T$ 
region, the two contributions to the crystal entropy variation stemming from Li$^{+}$ ionic diffusion and (BH)$_{12}^{-2}$ 
icosahedra reorientational motion have opposite signs hence $|\Delta S|$ decreases significantly. The identified 
anomalous lithium diffusion behaviour, however, ceases at $P \approx 0.6$~GPa since beyond that point $D_{\rm Li}$ 
decreases systematically upon increasing pressure (Supplementary Fig.1). 

\begin{table*}
\centering
\begin{tabular}{c c c c c c c c c}
\hline
\hline
$ $ & $ $ & $ $ & $ $ & $ $ & $ $ & $ $ & $ $ & $ $ \\
$ $ & \quad  $T$ \quad & \quad $ P $ \quad & \quad $|\Delta S|$ \quad & \quad $|\Delta T|$ \quad & \quad $|\Delta S|/P$ & \quad $|\Delta T|/P$ \quad & \quad ${\rm Material}$ \quad & \quad $ {\rm Reference} $ \\
$ $ \quad & \quad ${\rm (K)}$ \quad & \quad ${\rm (GPa)}$ \quad & \quad ${\rm (J K^{-1} kg^{-1})}$ \quad & \quad ${\rm (K)}$ \quad & \quad ${\rm (J K^{-1} kg^{-1} GPa^{-1})}$ \quad & \quad ${\rm (K~GPa^{-1})} $ \quad & \quad $ $ \quad & \quad $ $ \\
$ $ & $ $ & $ $ & $ $ & $ $ & $ $ & $ $ & $ $ & $ $ \\
\hline
$ $ & $ $ & $ $ & $ $ & $ $ & $ $ & $ $ & $ $ & $ $ \\
${\rm Ni_{51}Mn_{33}In_{16}}          $ \qquad & $ 330 $ & $ 0.25 $ & $ 41.0  $ & $ 4.0  $ & $ 164  $ & $ 16.0 $  & $ {\rm SMA} $ &  \cite{taulats15a}  \\
${\rm Fe_{49}Rh_{51}}                 $ \qquad & $ 310 $ & $ 0.11 $ & $ 12.5  $ & $ 8.1  $ & $ 114  $ & $ 73.6 $  & $ {\rm SMA} $ &  \cite{taulats15}  \\
${\rm (NH_{4})_{2}SO_{4}}             $ \qquad & $ 220 $ & $ 0.10 $ & $ 130.0 $ & $ 8.0  $ & $ 1300 $ & $ 80.0 $  & $ {\rm FE}  $ &  \cite{lloveras15} \\
${\rm [TPrA][Mn(dca)_{3}]}            $ \qquad & $ 330 $ & $ 0.01 $ & $ 30.5  $ & $ 4.1  $ & $ 3050 $ & $ 410.0$  & $ {\rm OIH} $ & \cite{bermudez17} \\
${\rm [FeL_{2}][BF_{4}]_{2}}          $ \qquad & $ 262 $ & $ 0.03 $ & $ 80.0  $ & $ 3.0  $ & $ 2667 $ & $ 100.0 $ & $ {\rm MC}  $ & \cite{vallone19} \\
${\rm (CH_{3})_{2}C(CH_{2}OH)_{2}}    $ \qquad & $ 320 $ & $ 0.52 $ & $ 510.0 $ & $ 45.0 $ & $ 981  $ & $ 86.5  $ & $ {\rm MC}  $ & \cite{li19,lloveras19} \\
${\rm AgI}                            $ \qquad & $ 400 $ & $ 0.25 $ & $ 62.0  $ & $ 36.0 $ & $ 248  $ & $ 144.0$  & $ {\rm FIC} $ & \cite{cazorla17a} \\
${\rm Li_{2}B_{12}H_{12}}             $ \qquad & $ 475 $ & $ 0.40 $ & $ 387.0 $ & $ 26.0 $ & $ 968  $ & $ 65.0 $  & $ {\rm FIC/MC} $ & $ {\rm This~work} $ \\
$ $ & $ $ & $ $ & $ $ & $ $ & $ $ & $ $ & $ $ & $ $ \\
\hline
\hline
\end{tabular}
\label{tab:mcperform}
 \caption{Materials presenting giant ($|\Delta S| \sim 10$~JK$^{-1}$kg$^{-1}$) and colossal 
	($\sim 100$~JK$^{-1}$kg$^{-1}$) barocaloric effects. $T$ represents working temperature, 
	$P$ applied pressure, $|\Delta S|$ isothermal entropy change, $|\Delta T|$ adiabatic 
	temperature change, $|\Delta T|/P$ barocaloric strength, ``SMA'' shape-memory alloy, ``FE'' 
	ferroelectric, ``OIH'' organic-inorganic hybrid perovskite, ``MC'' molecular crystal and 
	``FIC'' fast-ion conductor.}
\end{table*}

\section*{DISCUSSION}
To date, large BC effects have been experimentally measured for a number of shape-memory alloys \cite{taulats15a,taulats15}, 
polar compounds \cite{lloveras15}, organic-inorganic hybrid perovskites \cite{bermudez17,bermudez17b}, fluoride-based 
materials \cite{gorev10}, polymers \cite{rodriguez82}, the fast-ion conductor AgI \cite{cazorla17a} and molecular crystals 
\cite{li19,lloveras19,vallone19}. In Table~I, we compare the barocaloric performance predicted for bulk LBH with those of 
some representative barocaloric materials \cite{manosa17,moya14,cazorla19}. The isothermal entropy change induced in LBH by 
a moderate hydrostatic pressure of $0.4$~GPa, $387$~JK$^{-1}$kg$^{-1}$, is comparable in magnitude to the record $|\Delta S|$ 
that has been recently reported for the plastic crystal neopentylglycol by considering a similar pressure shift, 
$510$~JK$^{-1}$kg$^{-1}$ \cite{li19,lloveras19}. The rest of materials in Table~I present isothermal entropy changes 
that are appreciably smaller, made the exception of the polar crystal (NH$_{4}$)$_{2}$SO$_{4}$ which registers 
$130$~JK$^{-1}$kg$^{-1}$. As regards $|\Delta T|$, the clear contestants of LBH are the fast-ion conductor AgI ($36$~K) 
and again the plastic crystal (CH$_{3}$)$_{2}$C(CH$_{2}$OH)$_{2}$ ($45$~K). The reason for the smaller $|\Delta T|$ 
value estimated for LBH as compared to that of AgI is the significantly larger heat capacity of the former material, which 
results from a smaller molecular weight \cite{cazorla18}. In terms of the barocaloric strengths defined as ${\rm BSS} \equiv 
|\Delta S| / P$ and ${\rm BST} \equiv |\Delta T| / P$, LBH remains competitive with the best performers. For instance, the 
organic-inorganic hybrid perovskite [TPrA][Mn(dca)$_{3}$] displays the largest BSS and BST coefficients of all crystals, 
$\approx 3,000$~JK$^{-1}$kg$^{-1}$GPa$^{-1}$ and $\approx 400$~K~GPa$^{-1}$, respectively, while for bulk LBH we estimate 
$\approx 1,000$~JK$^{-1}$kg$^{-1}$GPa$^{-1}$ and $\approx 100$~K~GPa$^{-1}$. Meanwhile, the barocaloric strengths reported 
for the plastic crystal neopentylglycol are comparable in magnitude to those predicted for LBH, which hints at their 
common order-disorder phase-transition origin. 

As it was mentioned in the Introduction, the magnitude of the $|\Delta T|$ and $|\Delta S|$ shifts are not the only 
parameters that define the barocaloric performance of a material. The degree of reversibility of the involved $P$-induced 
phase transition, for instance, is another important barocaloric descriptor that provides information on the materials 
efficiency during successive pressure application/removal cycles. Specifically, the hysteresis of the transition makes 
the materials behaviour to depend on its cycling history and to increase the value of the external field that is 
required to bring the phase transition to completion \cite{lloveras20}. As a consequence, the barocaloric performance 
of a hysteretic material can be significantly worse than that of its ideal non-hysteretic counterpart. In order to 
quantify the degree of reversibility associated with the $\alpha \leftrightarrow \beta$ phase transition in LBH, we 
performed a series of long MD simulations ($\sim 2$~ns) in which the pressure (temperature) was kept fixed while the 
temperature (pressure) was varied steadily first from $425$ up to $625$~K (from $0.0$ up to $0.4$~GPa) and subsequently 
from $625$ back to $425$~K (from $0.4$ back to $0.0$~GPa). The results of such field-changing simulations indicate that 
the degree of reversibility of the order-disorder $\alpha \to \beta$ phase transition is quite acceptable (Supplementary 
Fig.2). For instance, by monitoring the variation of the system volume, we found that at zero pressure the difference 
between the transition temperatures observed during the heating and cooling stages was $\Delta T_{h} \equiv T_{\alpha 
\to \beta} - T_{\beta \to \alpha} \approx 50$~K (Supplementary Fig.2a). The size of $\Delta T_{h}$, however, increases 
noticeably at higher pressures ($\approx 100$~K at $0.4$~GPa). Meanwhile, at fixed temperature we found that the 
hysteresis of the phase transition as driven by pressure was practically null at $T = 550$~K ($\Delta P_{h} \approx 0$~GPa) 
and equal to $0.1$~GPa at $475$~K (Supplementary Fig.2b).  

Arguably the only weakness of bulk LBH in terms of barocaloric potential is that the critical temperature of the 
order-disorder $\alpha \to \beta$ phase transition is significantly higher than room temperature. However, this 
practical problem can be efficiently solved by means of doping and alloying strategies. In fact, recently it has 
been experimentally shown that carbon-doped LBH, LiCB$_{11}$H$_{12}$, presents a much lower $\alpha \to \beta$ 
transition temperature of $\approx 400$~K \cite{tang15}, and that the disordered $\beta$ phase is already stabilized 
at room temperature in Li(CB$_{9}$H$_{10}$)--Li(CB$_{11}$H$_{12}$) solid solutions \cite{kim19}. Moreover, the 
type of isosymmetric order-disorder phase transition underlying the exceptional barocaloric behaviour of LBH occurs 
also in analogous alkali-metal complex hydrides (A$_{2}$B$_{12}$H$_{12}$, A = Na, K, Cs) \cite{verdal11} and other  
earth-abundant and non-toxic materials like KHPO$_{4}$, NaAlSi$_{3}$O$_{8}$ and KNO$_{3}$ \cite{christy95}. Bulk 
KNO$_{3}$, for example, displays a staggering volume collapse of $\sim 10$\% for a room-temperature phase transformation 
induced by a modest pressure of $0.3$~GPa \cite{adams88}, which suggests great barocaloric potential as well.

In conclusion, we have predicted the existence of colossal barocaloric effects rendering isothermal entropy changes
of $\sim 100$~JK$^{-1}$kg$^{-1}$ and adiabatic temperature shifts of $\sim 10$~K in the complex hydride Li$_{2}$B$_{12}$H$_{12}$,
which are driven by moderate hydrostatic pressures of $\sim 0.1$~GPa. The phase transition underlying such colossal 
barocaloric effects is remarkable as it combines key ingredients of fast-ion conductors (i.e., ionic diffusion) 
and molecular crystals (i.e., reorientational motion), materials that individually have been proven to be excellent
barocaloric materials. This same type of isosymmetric order-disorder phase transition is likely to occur also in 
other economically affordable and innocuous compounds (e.g., Cs$_{2}$B$_{12}$H$_{12}$ and KNO$_{3}$), thus broadening
significantly the spectrum of caloric materials with commercial potential for solid-state cooling applications. We 
believe that our simulation study will stimulate experimental research on this new family of barocaloric materials, 
namely, alkali-metal complex hydrides, which are already known from other technological disciplines (e.g., hydrogen 
storage and electrochemical devices) and are routinely synthesized in the laboratory.

\section*{METHODS}
{\bf Classical molecular dynamics simulations.}~Molecular dynamics (MD) $(N, P, T)$ simulations were 
performed with the LAMMPS code \cite{lammps}. The pressure and temperature in the system were kept
fluctuating around a set-point value by using thermostatting and barostatting techniques in which some
dynamic variables are coupled to the particle velocities and simulation box dimensions. The interactions
between atoms were modeled with the harmonic Coulomb-Buckingham interatomic potential reported in work 
\cite{sau19}, the details of which are provided in the Supplementary Methods. The employed interatomic 
potential reproduces satisfactorily the vibrational spectra, structure and lithium diffusion coefficients 
of bulk LBH \cite{sau19} (Supplementary Discussion). We employed simulation boxes containing $6656$ atoms 
and applied periodic boundary conditions along the three Cartesian directions. Newton's equations of motion 
were integrated using the customary Verlet's algorithm with a time-step length of $0.5$~fs. The typical 
duration of a MD $(N, P, T)$ run was of $1$~ns. A particle-particle particle-mesh $k$-space solver was 
used to compute long-range van der Waals and Coulomb interactions beyond a cut-off distance of $10$~\AA~ 
at each time step. 
\\

{\bf Density functional theory calculations.}~First-principles calculations based on density functional theory (DFT) 
\cite{cazorla17} were performed to analyse the energy, structural, vibrational, and ionic transport properties of 
Li$_{2}$B$_{12}$H$_{12}$. We performed these calculations with the VASP software \cite{vasp} by following the generalized 
gradient approximation to the exchange-correlation energy due to Perdew \emph{et al.} \cite{pbe96}. The projector 
augmented-wave method was used to represent the ionic cores \cite{bloch94}, and the electronic states $1s$-$2s$  
Li, $1s$-$2s$-$2p$ B and $1s$ H were considered as valence. Wave functions were represented in a plane-wave basis 
set truncated at $650$~eV. By using these parameters and dense ${\bf k}$-point grids for Brillouin zone integration, 
the resulting energies were converged to within $1$~meV per formula unit. In the geometry relaxations, a tolerance 
of $0.01$~eV$\cdot$\AA$^{-1}$ was imposed on the atomic forces. 

\emph{Ab initio} molecular dynamics (AIMD) simulations based on DFT were carried out to assess the reliability of 
the interatomic potential model employed in the classical molecular dynamics simulations (Supplementary Fig.3 and
Supplementary Discussion). The AIMD simulations were performed in the canonical $(N,V,T)$ ensemble considering 
constant number of particles, volume and temperature. The constrained volumes were equal to the equilibrium volumes 
determined at zero temperature, thus we neglected possible thermal expansion effects. Nevertheless, in view of 
previous first-principles work \cite{cazorla19b}, it is reasonable to expect that thermal expansion effects do not 
affect significantly the estimation of lithium diffusion coefficients at the considered temperatures. The temperature 
in the AIMD simulations was kept fluctuating around a set-point value by using Nose-Hoover thermostats. A large 
simulation box containing $832$ atoms was employed in all the simulations, and periodic boundary conditions were 
applied along the three Cartesian directions. Newton's equations of motion were integrated by using the customary 
Verlet's algorithm and a time-step length of $\delta t = 10^{-3}$~ps. $\Gamma$-point sampling for integration within 
the first Brillouin zone was employed in all the AIMD simulations. The AIMD simulations comprised long simulation 
times of $\sim 100$~ps.
\\

{\bf Estimation of key quantities.}~The mean square displacement of lithium ions was estimated with the 
formula \cite{cazorla19b}:
\begin{eqnarray}
{\rm MSD_{\rm Li}}(\tau) & = & \frac{1}{N_{\rm ion} \left( N_{\rm step} - n_{\tau} \right)} \times \\ \nonumber
                &   & \sum_{i=1}^{N_{\rm ion}} \sum_{j=1}^{N_{\rm step} - n_{\tau}} | {\bf r}_{i} (t_{j} + \tau) - {\bf r}_{i} (t_{j}) |^{2}~, 
\label{eq1}
\end{eqnarray}
where ${\bf r}_{i}(t_{j})$ is the position of the migrating ion $i$ at time $t_{j}$ ($= j \cdot \delta t$), $\tau$
represents a lag time, $n_{\tau} = \tau / \delta t$, $N_{\rm ion}$ is the total number of mobile ions, and $N_{\rm step}$
the total number of time steps. The maximum $n_{\tau}$ was chosen equal to $N_{\rm step}/2$, hence we could accumulate 
enough statistics to reduce significantly the fluctuations in ${\rm MSD_{\rm Li}}(\tau)$ at large $\tau$'s. The diffusion 
coefficient of lithium ions then was obtained with the Einstein relation:
\begin{equation}
D_{\rm Li} =  \lim_{\tau \to \infty} \frac{{\rm MSD_{Li}}(\tau)}{6\tau}~,  
\label{eq2}     
\end{equation}
by performing linear fits to the averaged ${\rm MSD_{Li}}$ values calculated at long $\tau$.

The angular autocorrelation function of the closoborane (BH)$_{12}^{2-}$ icosahedra was estimated according to the 
expression \cite{sau19}:
\begin{equation}
	\phi_{\rm B_{12}H_{12}} (\tau) = \langle \hat{{\bf r}} (t) \cdot \hat{{\bf r}} (t + \tau) \rangle~,
\label{eq3}
\end{equation}	
where $\hat{{\bf r}}$ is a unitary vector connecting the center of mass of each closoborane unit with one of its edges 
and $\langle \cdots \rangle$ denotes thermal average considering all the closoborane icosahedra. This autocorrelation function 
typically decays as $\propto \exp{[-\lambda_{\rm B_{12}H_{12}} \cdot \tau]}$, where the parameter $\lambda_{\rm B_{12}H_{12}}$ 
represents a characteristic reorientational frequency. When the (BH)$_{12}^{2-}$ reorientational motion is significant, 
that is, $\lambda_{\rm B_{12}H_{12}}$ is large, the $\phi_{\rm B_{12}H_{12}}$ function decreases rapidly to zero with time.
 
Isothermal entropy changes associated with the barocaloric effect were estimated with the formula \cite{moya14,cazorla19}:
\begin{eqnarray}
\Delta S (P, T) =  - \int_{0}^{P} \left(\frac{\partial V}{\partial T}\right)_{P'} dP'~,
\label{eq4}
\end{eqnarray}
where $P$ represents the maximum applied hydrostatic pressure and $V$ the volume of the system. Likewise, the 
accompanying adiabatic temperature shift was calculated as:
\begin{equation}
\Delta T (P, T) = \int_{0}^{P} \frac{T}{C_{P'}(T)} \cdot \left(\frac{\partial V}{\partial T}\right)_{P'} dP'~, 
\label{eq5}
\end{equation}
where $C_{P}(T) = \left( \frac{dU}{dT} \right)_{P}$ is the heat capacity of the crystal obtained at constant pressure 
and temperature conditions.

In order to accurately compute the $\Delta S (P, T)$ and $\Delta T (P, T)$ shifts induced by pressure, we calculated 
the corresponding volumes and heat capacities over dense grids of $(P, T)$ points spaced by $\delta P = 0.1$~GPa and 
$\delta T = 25$~K. Spline interpolations were subsequently applied to the calculated sets of points, which allowed for 
accurate determination of $\left( \partial V / \partial T \right)_{P}$ and heat capacities. The $\Delta S$ and $\Delta T$ 
values appearing in Fig.\ref{fig2}d--e were obtained by numerically integrating those spline functions with respect to 
pressure.

\section*{DATA AVAILABILITY}
The data that support the findings of this study are available from the corresponding author (C.C.) upon
reasonable request.

\section*{ACKNOWLEDGEMENTS}
C. C. acknowledges support from the Spanish Ministry of Science, Innovation and Universities under the ``Ram\'on y Cajal'' 
fellowship RYC2018-024947-I. D. E. acknowledges support from the Spanish Ministry of Science, Innovation and Universities 
under the Grant PID2019-106383GB-C41 and the Generalitat Valenciana under the Grant Prometeo/2018/123 (EFIMAT). Computational 
resources and technical assistance were provided by the Informatics Service of the University of Valencia through the Tirant 
III cluster and the Center for Computational Materials Science of the Institute for Materials Research, Tohoku University 
(MAterial science Supercomputing system for Advanced MUltiscale simulations towards NExt-generation-Institute of Material 
Research) (Project No-19S0010).

\section*{AUTHOR CONTRIBUTIONS}
K.S. and C.C. conceived the study and planned the research. K.S. performed the molecular dynamics
simulations and C.C. the first-principles calculations and barocaloric analysis. Results were 
discussed by all the authors. All the authors participated in the writing of the manuscript.
\\

\section*{ADDITIONAL INFORMATION}
Supplementary information is available in the online version of the paper.
\\

\section*{COMPETING INTERESTS}
The authors declare no competing interests.

\end{document}